\documentclass{desyproc}
\newcommand {\pom} {I\!\! P}
\begin{document}
\title{Phenomenology of single and double diffraction dissociation}

\author{{\slshape Konstantin Goulianos}\\[1ex]
The Rockefeller University, 1230 York Avenue, New York, NY 10065, USA}

\contribID{smith\_joe}


\acronym{EDS'09} 

\maketitle

\begin{abstract}
Predictions of the gap-probability renormalization model for single and double diffraction dissociation cross sections in proton-proton collisions at the LHC are presented and compared with recent CMS measurements.
\end{abstract}

\section{Introduction}\label{sec:intro}
Measurements at the {\sc lhc} have shown that there are sizable disagreements among Monte Carlo {\sc (mc)} implementations of ``soft'' processes based on cross sections proposed by various physics models, and that it is not possible to reliably predict all such processes, or even all aspects of a given process, using a single model~\cite{ref:d2012_pheno,D2012_talks_vs_models,ref:dis13_pheno}.
In the {\sc cdf} studies of diffraction at the Tevatron, all processes are well modeled by the {\sc mbr} (Minimum Bias Rockefeller) {\sc mc} simulation, which is a stand-alone simulation based on a unitarized Regge-theory model, {\sc renorm}~\cite{RENORM}, employing inclusive nucleon parton distribution functions ({\sc pdf}'s) and {\sc qcd} color factors. The {\sc renorm} model was updated in a presentation at {\sc eds-2009}~\cite{EDS2009_total} to include a unique unitarization prescription for predicting the total $pp$ cross section at high energies, and that update has been included as an {\sc mbr} option for simulating diffractive processes in {\sc pythia8} since version  {\sc pythia8}.165~\cite{PYTHIA8.165}, to be referred here-forth as {\sc pythia8-mbr}. In this paper, we briefly review the cross sections~\cite{MBR_note} implemented in this option of {\sc pythia8} and compare the {\sc sd} and {\sc dd} predictions with {\sc lhc} measurements.     


\section{Cross sections}
The following diffraction dissociation processes are considered in {\sc pythia8-mbr}:
\begin{eqnarray}
\hbox{\sc sd}& pp\rightarrow Xp&{\rm Single\:Diffraction\;(or\;Single\;Dissociation)},\\
{\rm or}&pp\rightarrow pY&{\rm (the\;other\;proton\;survives)}\nonumber\\
\hbox{\sc dd}&pp\rightarrow XY&{\rm Double\;Diffraction\;(or\;Double\;Dissociation)},\\
\hbox{{\sc cd} (or {\sc dpe})}&pp\rightarrow pXp&{\rm Central\;Diffraction\;(or\;Double\;Pomeron\;Exchange)}.
\label{eqn:processes}
\end{eqnarray}

The {\sc renorm} predictions are expressed as unitarized Regge-theory formulas, in which the unitarization is achieved by a renormalization scheme where the Pomeron ($\pom$) flux is interpreted as the probability for forming a diffractive (non-exponentially suppressed) rapidity gap and thereby its integral over all phase space saturates at the energy where it reaches unity. Differential cross sections are expressed in terms of the $\pom$-trajectory,  $\alpha(t)=1+\epsilon +\alpha't = 1.104 + 0.25~\mbox({\rm GeV}^{-2})\cdot t$, the $\pom$-$p$ coupling, $\beta(t)$, and the ratio of the triple-$\pom$ to the $\pom$-$p$ couplings, $\kappa \equiv g(t)/\beta(0)$. For large rapidity gaps, $\Delta y\geq 3$, for which $\pom$-exchange dominates, the cross sections may be written as,  
\begin{eqnarray}
\frac{d^2\sigma_{SD}}{dtd\Delta y} & = & \frac{1}{N_{\rm gap}(s)} \left[ \frac{~ ~ \beta^2(t)}{16\pi}e^{2[\alpha(t)-1]\Delta y}\right] \cdot \left\{ \kappa \beta^2(0) \left( \frac{s'}{s_{0}}\right)^{\epsilon}\right\}, \label{eqSD}\\
\frac{d^3\sigma_{DD}}{dtd\Delta y dy_0} & = & \frac{1}{N_{\rm gap}(s)} \left[ \frac{\kappa\beta^2(0)}{16\pi}e^{2[\alpha(t)-1]\Delta y}\right] \cdot \left\{ \kappa \beta^2(0) \left( \frac{s'}{s_{0}}\right)^{\epsilon}\right\}, \label{eqDD}\\
\frac{d^4\sigma_{DPE}}{dt_1dt_2d\Delta y dy_c} & = & \frac{1}{N_{\rm gap}(s)} \left[\Pi_i\left[ \frac{\beta^2(t_i)}{16\pi}e^{2[\alpha(t_i)-1]\Delta y_i}\right]\right] \cdot \kappa \left\{ \kappa \beta^2(0) \left( \frac{s'}{s_{0}}\right)^{\epsilon}\right\}, \label{eqCD}
\end{eqnarray}
where $t$ is the 4-momentum-transfer squared at the proton vertex, $\Delta y$ the rapidity-gap width, and $y_0$  the center of the rapidity gap. In Eq.~(\ref{eqCD}), the subscript $i=1, 2$ enumerates Pomerons in a {\sc dpe} event, $\Delta y=\Delta y_1 + \Delta y_2$ is the total rapidity gap (sum of two gaps) in the event, and $y_c$ is the center in $\eta$ of the centrally-produced hadronic system.



\section{Results}
In this section, we present as examples of the predictive power of the {\sc renorm} model some results reported by the {\sc totem}, {\sc cms}, and {\sc alice} collaborations for $pp$ collisions at $\sqrt s=7$~TeV, which can be directly compared with {\sc renorm} formulas without using the {\sc pythia8-mbr} simulation. 

Another example of the predictive power of {\sc renorm} is shown in Fig.~2, which displays the total {\sc sd} (left) and total {\sc dd} (right) cross sections for $\xi<0.05$, after extrapolation into the low mass region from the measured {\sc cms} cross sections at higher mass regions, presented in~\cite{ref:ciesielski}, using {\sc renorm}.   
\begin{center}
\begin{figure}[!ht]
\begin{center}
\includegraphics[width=0.47\textwidth]{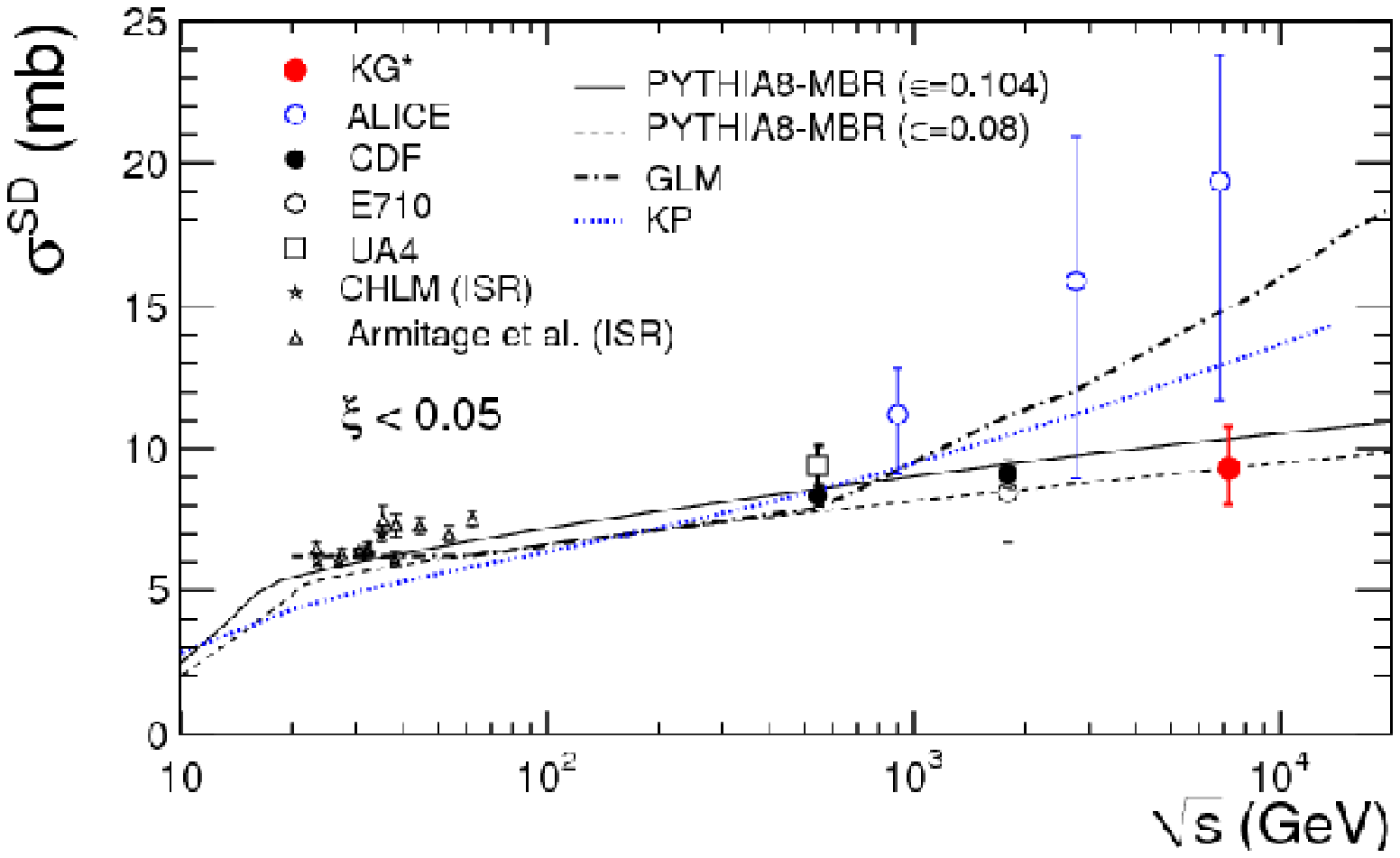}
\includegraphics[width=0.47\textwidth]{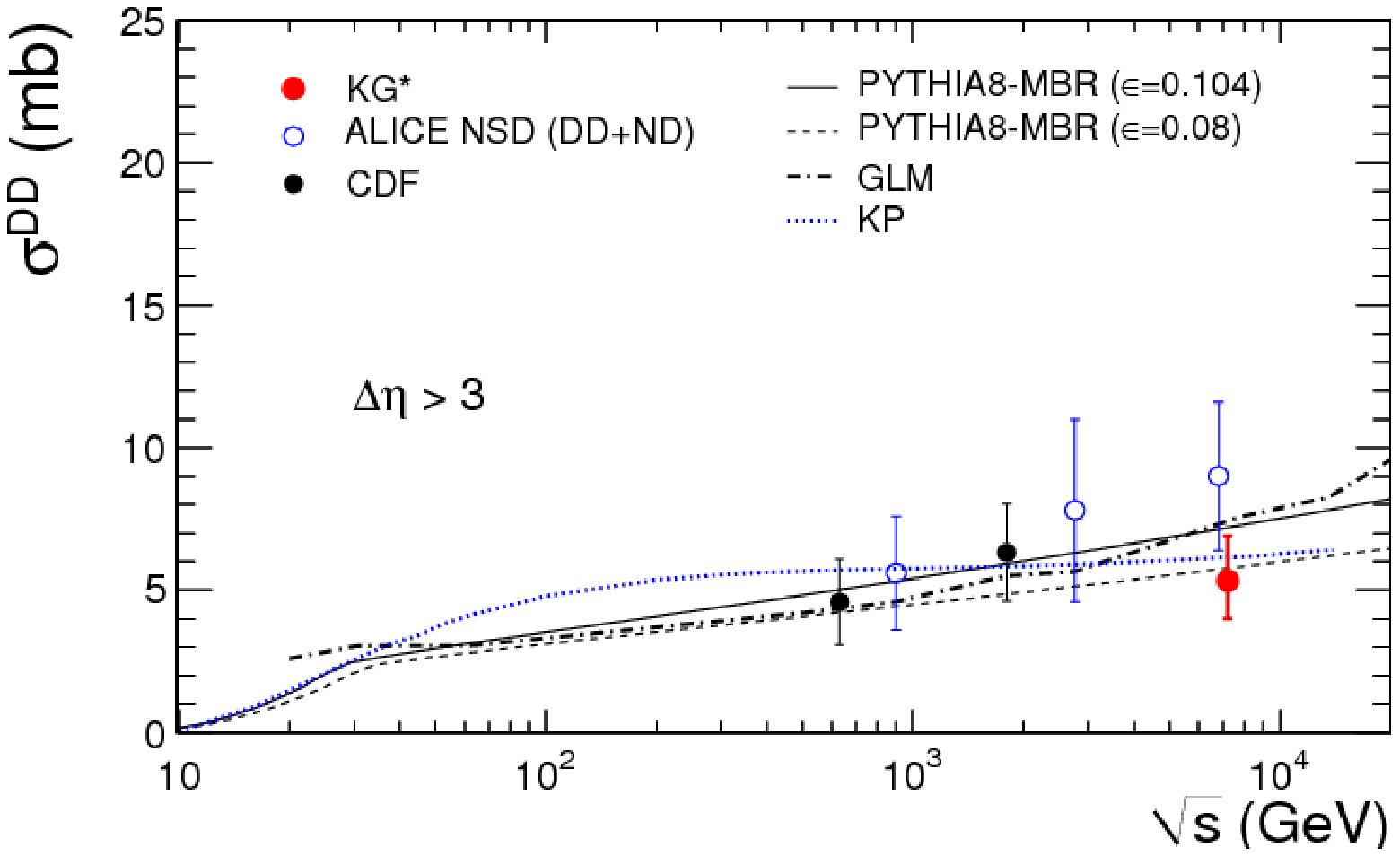}\\
KG*: this ``data'' point was obtained after extrapolation into the unmeasured low mass region(s) from the measured {\sc cms} cross sections~\cite{ref:ciesielski} using the {\sc mbr} model.
\caption{Measured {\sc sd} (left) and {\sc dd} (right) cross sections for $\xi<0.05$ compared with theoretical predictions; the model embedded in {\sc pythia8-mbr} provides a good description of all data.}
\label{fig:fig2}
\end{center}\end{figure}
\end{center}

\section{Summary\label{sec:conclusion}}
 Pre-{\sc lhc} predictions for the {\sc sd} and {\sc dd} cross sections at high energies, based on the {\sc renorm} special parton-model approach to diffraction, employing inclusive proton parton distribution functions and {\sc qcd} color factors have been reviewed. The predictions of the model are in good agreement with the {\sc cms} results presented at this conference~\cite{KG:cmsdiff}.

\section{Acknowledgments}
I would like to thank the Office of Science of the Department of Energy for supporting the Rockefeller experimental diffraction physics programs at Fermilab and {\sc lhc} on which this research is anchored.


\begin{footnotesize}

\end{footnotesize}
\end{document}